\documentclass[manuscript]{aastex631}
\shorttitle{AASTeX v6.3.1 Sample article}
\shortauthors{Lu et al.}
\graphicspath{{./}{figures/}}

\begin{document}

\title{Spatially Resolved Moving Radio Burst in Association with an EUV Wave}

\correspondingauthor{Lei Lu}
\email{leilu@pmo.ac.cn}
\correspondingauthor{Li Feng}
\email{lfeng@pmo.ac.cn}

\author[0000-0002-3032-6066]{Lei Lu}
\altaffiliation{Chang Yu-Che Fellow of Purple Mountain Observatory}
\affiliation{Key Laboratory of Dark Matter and Space Astronomy, Purple Mountain Observatory, Chinese Academy of Sciences, Nanjing 210023, China}
\affiliation{CAS Key Laboratory of Solar Activity, National Astronomical Observatories, Beijing 100012, China}

\author[0000-0003-4655-6939]{Li Feng}
\affiliation{Key Laboratory of Dark Matter and Space Astronomy, Purple Mountain Observatory, Chinese Academy of Sciences, Nanjing 210023, China}

\author[0000-0001-9979-4178]{Weiqun Gan}
\affiliation{Key Laboratory of Dark Matter and Space Astronomy, Purple Mountain Observatory, Chinese Academy of Sciences, Nanjing 210023, China}

%
%
%
%
%



\begin{abstract}
Coronal mass ejections (CMEs) are large clouds of magnetized plasma ejected from the Sun, and are often associated with acceleration of electrons that can result in radio emission via various mechanisms.
However, the underlying mechanism relating the CMEs and particle acceleration still remains a subject of heated debate.
Here, we report multi-instrument radio and extreme ultraviolet (EUV) imaging of a solar eruption event on 24 September
2011. We determine the emission mechanism of a moving radio burst,  identify its three-dimensional (3D) location with respect to a rapidly expanding EUV wave, and find evidence for CME shocks that produce quasiperiodic acceleration of electron beams.  
\end{abstract}



\section{Introduction} \label{sec:intro}


Coronal mass ejections (CMEs) are large eruptions of magnetized  plasma from the low solar corona into  the interplanetary space, usually observed as a three-part structure in white-light coronagraphs, i.e., a dark cavity with a bright core embedded inside a bright compression front \citep{Webb2012}. 
The dark cavity is supposed to be the cross section of an expanding flux rope \citep{Chen1997}.
Fast CMEs can drive plasma shocks that are capable of accelerating electrons up to relativistic speeds,  generating bursts of radiation at radio wavelengths through the plasma emission mechanism, such as type II radio burst \citep{Klassen2002,Grechnev2011,Ying2018,Feng2020}.
Fine-structured bursts of radiation drifting towards both lower and higher frequencies, called herringbones, can sometimes be identified in dynamic spectra of type II bursts. These have been considered as signatures of individual electron beams accelerated by CME-driven shocks \citep{Mann2005}.

In some cases, CMEs are also  accompanied by broadband continuum radiation at decimetric and metric wavelengths, named as type IV radio burst. Over the years, type IV radio bursts have been subcategorized  into  moving Type IV bursts and stationary Type IV bursts  according to the motion of their source regions \citep{McLean1985,Pick2008}.  
Of particular interest are the moving type IV radio bursts, which  are commonly believed to be produced by energetic electrons trapped in CMEs via various emission mechanisms \citep{Carley2020}. 
So far, four emission mechanisms have been suggested to interpret the moving type IV radiations, including  coherent plasma emission \citep{Duncan1981,Hariharan2016,Vasanth2019} and electron cyclotron maser emission \citep{Melrose1982,Liu2018}, as well as incoherent synchrotron and gyrosynchrotron emission \citep{Kai1969,Carley2017}. 
Therefore, if the emission mechanism can be unambiguously determined, the type moving IV bursts can provide powerful diagnostics of plasma conditions inside CMEs, such as electron density,  characteristics of the electron energy distribution, or magnetic field strength \citep{Vourlidas2004,Pick2004,Carley2020}. 

The combination of radio and EUV images is helpful to identify the sites of electron acceleration and provide important information of the source region \citep{Chen2014,Vasanth2019,Morosan2020}.
In this letter, making use of advanced radio and EUV instruments, we present a spatially resolved moving radio burst that propagated closely in association with a bright EUV front.
The moving radio burst was found to be coincident with CME-driven shocks, but not show clear type II or herringbone structures. It is  superimposed on the type IV emission at high frequencies, but  behaves quite differently from the type IV burst at low frequencies.
We aim to determine the possible emission mechanism behind the moving radio burst and study its relations with the observed EUV phenomena.


\section{The data and method}

The propagation of the bright coronal structure is traced in 193 {\AA} images obtained by  the Atmospheric Imaging Assembly (AIA) \citep{Lemen2012} aboard Solar Dynamic Observatory (SDO). AIA  routinely takes full-disk images of the Sun and the low corona (up to 0.5 R$_{\odot}$ above the solar limb) in seven EUV wavebands (94 {\AA}, 131 {\AA}, 171 {\AA}, 193 {\AA}, 211 {\AA}, 304 {\AA}, and 335 {\AA})  and two UV wavebands (1600 {\AA} and 1700 {\AA}).
The EUV images are taken with a time cadence of 12 s and a spatial resolution of 1.2 arcsec.

The radio spectral data were obtained by  the ground-based  CALLISTO spectrometers \citep{Benz2009}, the Nancay Decameter Array (NDA) \citep{Lamy2017}, and the space-based WIND/WAVES instrument \citep{Bougeret1995}.
The CALLISTO spectrometers form a worldwide spectrometer network (e-CALLISTO) that is able to observe the solar radio spectrum from 10 MHz to 870 MHz with a time resolution up to 0.25 seconds for 24 hours per day.
The NDA routinely observes Jupiter and the Sun (at best $\sim$8 h per day) in 10$-$80 MHz frequency range with a time cadence of 1 second from Earth.  The WIND/WAVES provides a 24-hour measurement (from space) of low frequency (0.02$-$13.825 MHz) radio emissions of the Sun with a time cadence of 1 minute.

Positions of the radio sources  are measured by the Nancay Radioheliograph (NRH) \citep{Kerdraon1997}. NRH routinely observes the Sun for about 7 hours per day at six (before May 2008) then ten frequencies (after May 2008) between 150 and 450 MHz. 
It can provide the flux density as well as the brightness temperature data for Stokes I and  Stokes V components, respectively, with the highest time resolution up to  125 ms. Here we use the integrated data with a 1-second resolution for a higher signal-to-noise ratio.

The temporal evolution of radio emission was compared with hard X-ray radiations measured by the Reuven Ramaty High Energy Solar Spectroscopic Imager (RHESSI) \citep{Lin2002}. RHESSI is capable of providing high resolution imaging and spectroscopy of solar flares    from soft X-rays (3 keV) to gamma-rays (17 MeV).
The radio observations in combination with X-ray measurements provide the most direct  diagnostics of energetic electrons in the solar atmosphere and low corona.

Finally, EUV images at 195 {\AA}, taken by the Extreme Ultraviolet Imager (EUVI) aboard the STEREO-B spacecraft from another perspective, are used to provide nearly a top view of the active region.

\section{Results and discussions} \label{sec:obs}

\begin{figure*}[!h]
\centering
\includegraphics[scale=1.]{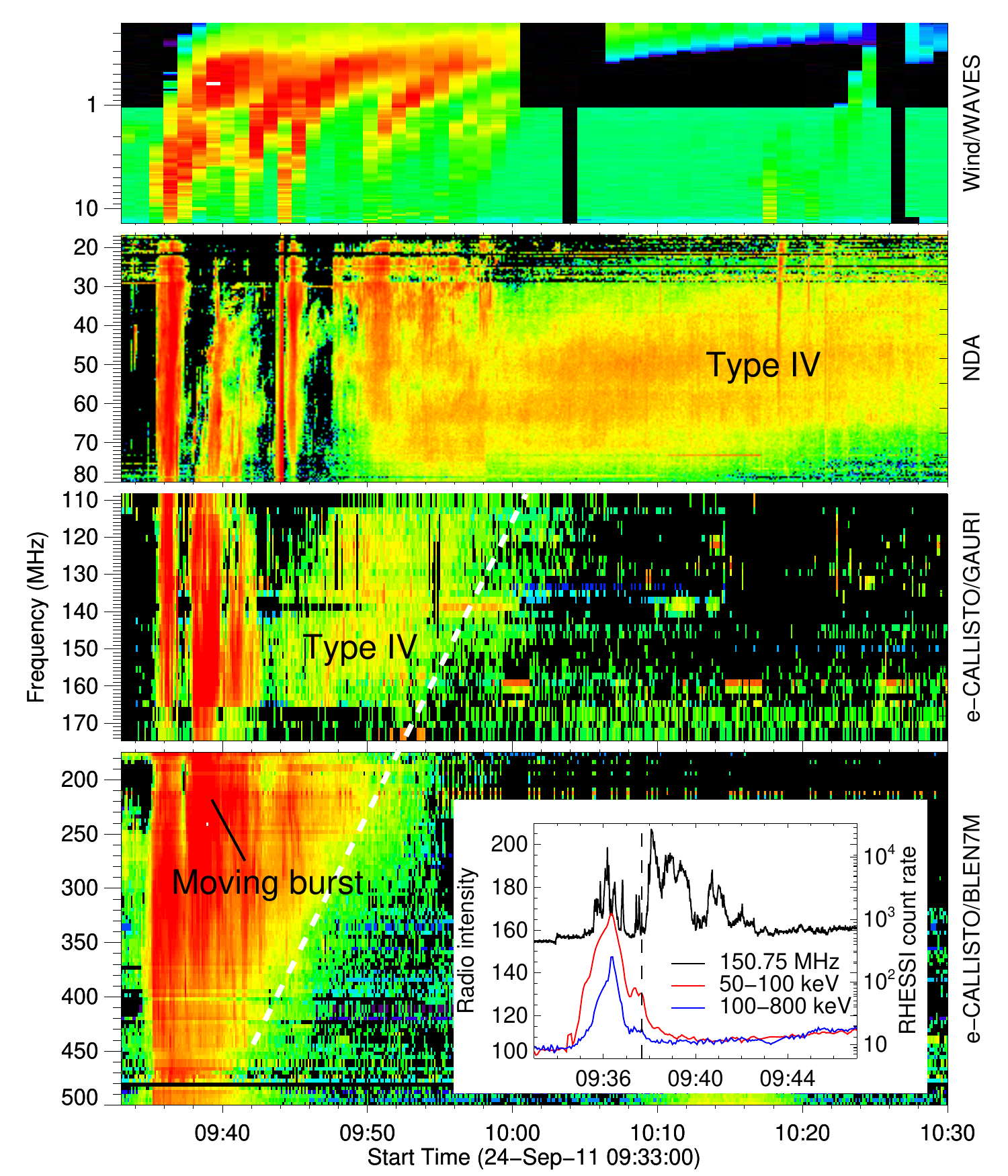}
\caption{Radio dynamic spectrum from WIND/WAVES, NDA, and e-CALLISTO spectrometers, showing the solar radio bursts on 24 September 2011. The white dashed line illustrates the type IV radio burst and the black line refers to the moving radio burst under study. The insert on the bottom panel shows the time profiles of the radio emission at 150.75 MHz and the hard X-ray emission above 50 keV.
\label{fig:dynamic-spectra}}
\end{figure*}

\begin{figure*}
\centering
\includegraphics[scale=.9]{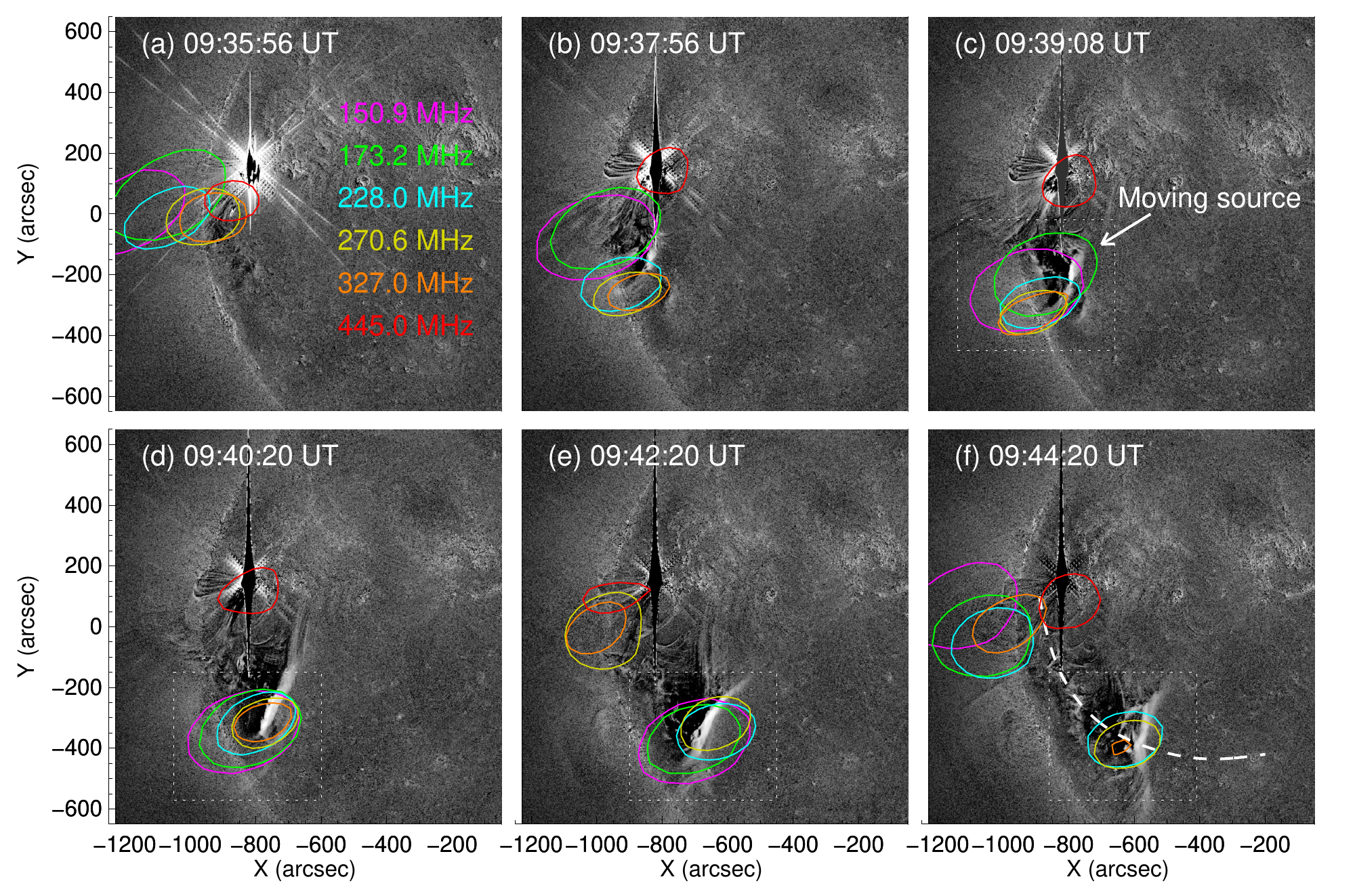}
\caption{ Time series of radio contours (70\% of the image maximum) at six NRH frequencies,overlaid on concurrent AIA 193 {\AA} running difference images.
(a) Source region of the type III radio burst. (b)-(f) Source region of the moving radio burst, which is co-spatial with an EUV bright front. The dashed box outlines the moving radio burst. The dashed line in panel (f) show the propagation direction of the EUV bright front.  An animation showing the co-spatial propagation of the radio source and the EUV bright front is available. It covers a duration of $\sim$ 6 minutes from 09:35 UT to 09:51 UT on 24 September 2011.The entire movie runs for $\sim$ 4 s.
\label{fig:bfront}}
\end{figure*}

The GOES X1.9 flare of 24 September 2011 occurred in NOAA active region 11302 near the eastern limb of the solar disc, and was followed by a series of radio bursts (see the dynamic spectrum in Fig.~\ref{fig:dynamic-spectra}).  From about 09:35 UT, shortly after the appearance of the flare, a type-III-like radio burst was observed. 
The type III burst is well correlated with the impulsive enhancement of the hard X-ray ($>$ 50 keV) emission measured by RHESSI (see insert in Fig. \ref{fig:dynamic-spectra}) and extends below 14 MHz, implying that the electrons were accelerated by the solar flare and injected into the interplanetary space at heights much larger than 1 R$_\odot$ above the solar surface.
Following the type III burst, a broadband type IV burst was observed to drift from high to low frequencies with time (indicated by dashed line in Fig.~\ref{fig:dynamic-spectra}), presumably ascribed to energetic electrons trapped in CMEs. The type IV emission lasts more than one hour and diminished sharply at about 30 MHz.
Of particular interest is the much stronger radio burst that started from about 09:37:40 UT.  At high frequencies, the radio burst was superimposed on the type IV burst while at low frequencies (below 14 MHz) it seems to correlate with  interplanetary type III radio bursts. Since the radio burst shows a clear movement in Fig.~\ref{fig:bfront}, we thus refer to the radio burst as moving radio burst throughout this paper.
Different from the earlier type III burst, the moving radio burst shows no relation with the hard X-ray emission (see insert in Fig. \ref{fig:dynamic-spectra}), implying a different physical origin.

\begin{figure*}
\centering
\includegraphics[scale=.9]{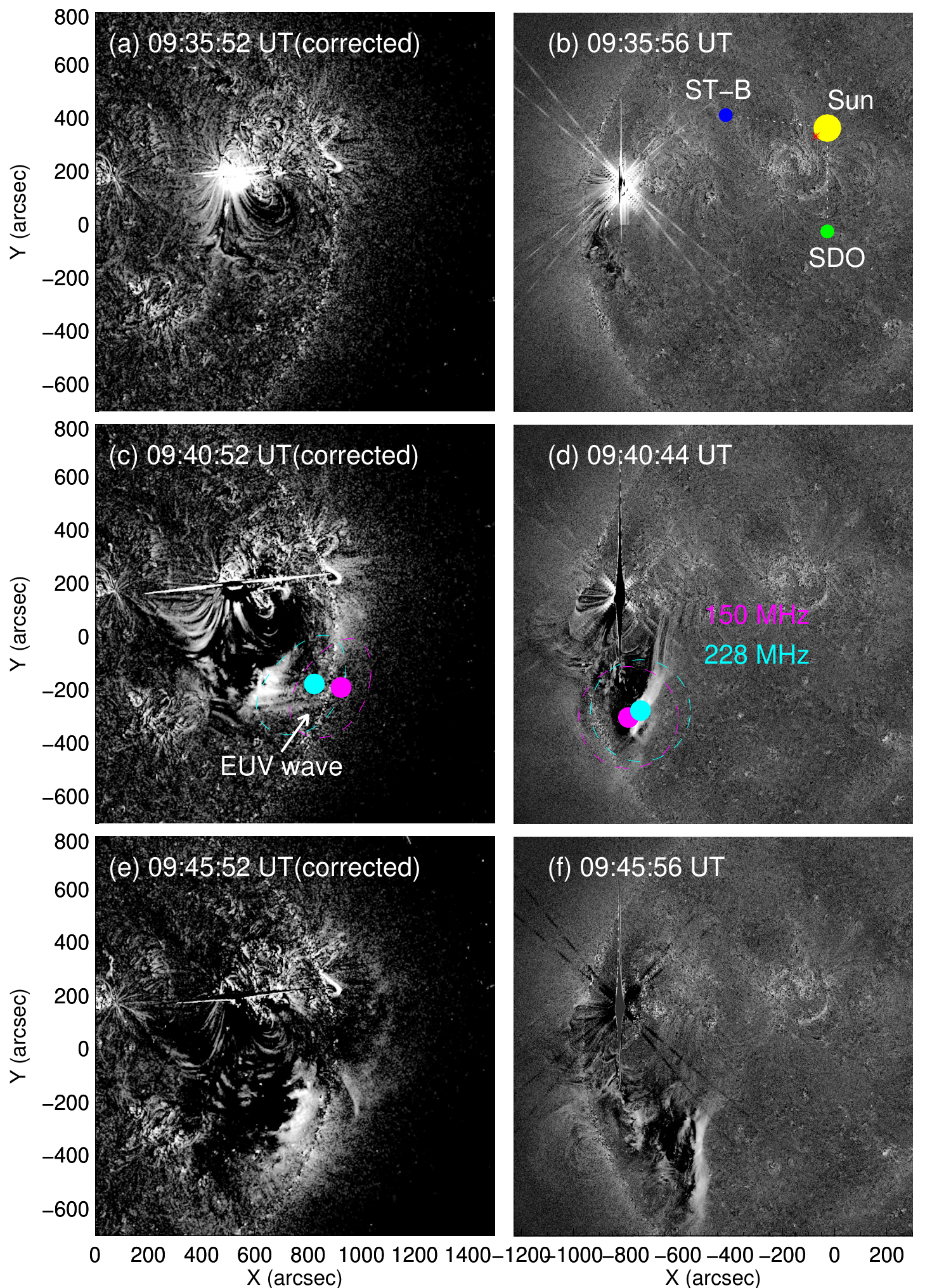}
\caption{ Comparison of images as viewed from STEREO-B/EUVI in 195 {\AA} (left) and SDO/AIA in 193 {\AA} (right) at three different times (from top to bottom). The observation time of STEREO-B has been corrected by supposing that the spacecraft are located at 1 AU (Astronomical Unit) from the Sun.
The insert in panel (b) shows the locations of SDO and STEREO-B in the ecliptic plane relative to the Sun.
The enlarger dots and dashed circles in panel (d) represent the radio emission centroids and their uncertainties ($\pm 0.2~R_{\odot}$) at 150 MHz (magenta) and 228 MHz (cyan), respectively, and their projections onto the STEREO-B/EUVI images are presented in panel (c).
The arrow points to an EUV wave with a coronal dimming behind. 
\label{fig:multiview}}
\end{figure*}

The source regions of the radio bursts are shown in Fig.~\ref{fig:bfront}, where  coloured contours  (70\% of the maximum intensity) at six NRH frequencies between 150 and 445 MHz are overlaid on concurrent AIA 193 {\AA} running difference images to provide information on the context in which the radio emission occurred. 
Fig.~\ref{fig:bfront}a shows the type III burst, during which the radio emission mainly arises from coronal regions extending above the flaring site, indicating the acceleration of energetic electrons by the solar flare \citep{LinJ2000}. 
Then after about 09:37:40 UT, the moving radio burst starts, together with a bright coronal front that propagates in southwest direction across the solar disc (Fig.~\ref{fig:bfront}b-f). 
Comparison of images as viewed from STEREO-B/EUVI in 195 {\AA} and SDO/AIA in 193 {\AA} (Fig.~\ref{fig:multiview}) shows that the bright front is probably correspond to the southwest flank of a rapidly expanding EUV wave \citep{Shen2012}.
The source region of the moving radio burst was found to be highly co-spatial with the bright front, implying a strong causal relationship between these two phenomena.  
A movie that shows the co-propagation of the radio sources and the bright front can be found in the online journal.  

\begin{figure*}
\centering
\includegraphics[scale=.9]{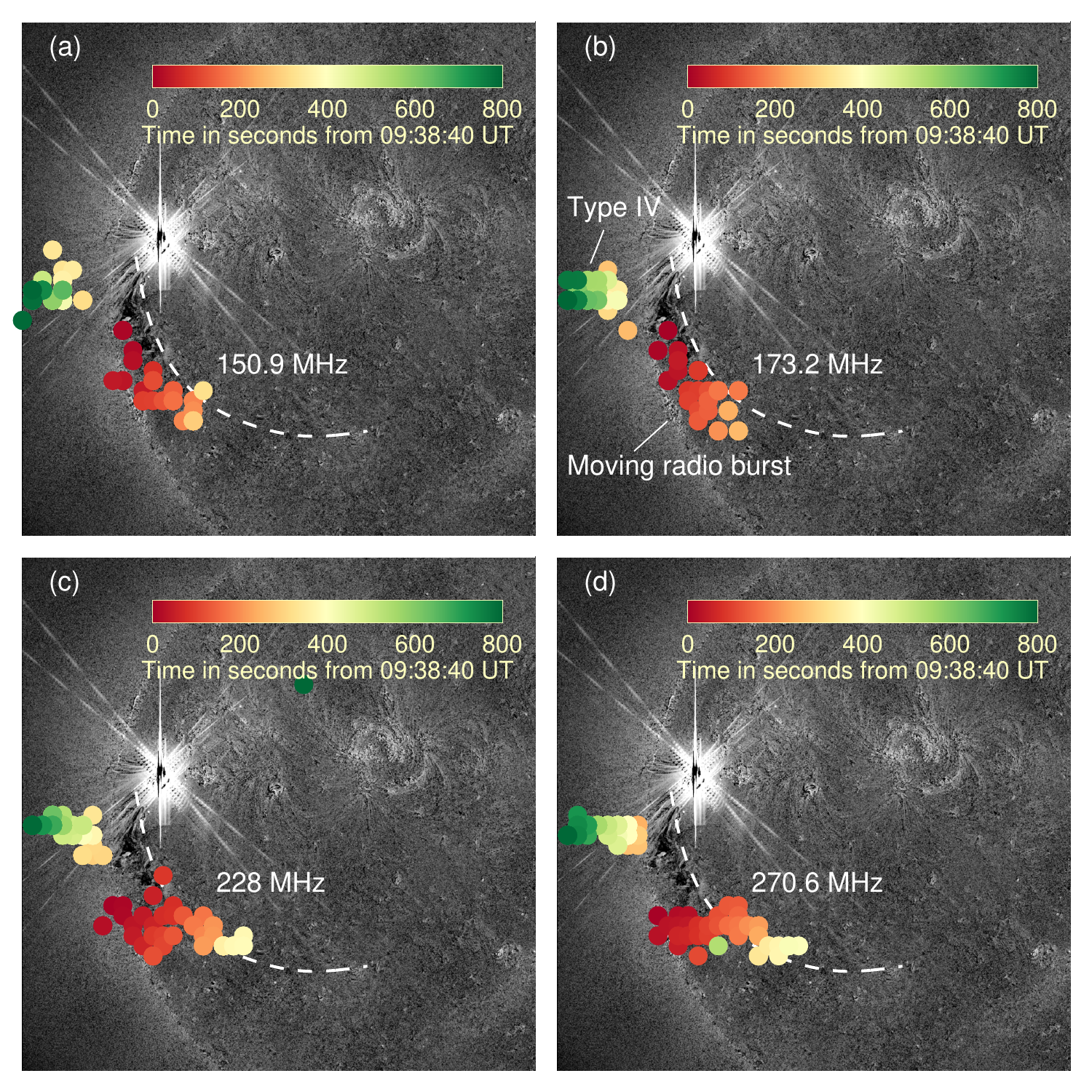}
\caption{Radio emission centroids coloured with progressing time, showing the propagation direction of the radio bursts on the POS images at AIA 193 {\AA}. The different panels correspond to the radio emission at different frequencies. The dashed line has the same meaning as that in Fig.~\ref{fig:bfront}f.
\label{fig:nrh_source_temevo}}
\end{figure*}

Fig.~\ref{fig:nrh_source_temevo} presents the radio centroids coloured with progressing times at four individual NRH frequencies, which shows the propagation direction of the radio bursts on the plane-of-sky (POS) images at 193 {\AA} from SDO/AIA. 
During the early time, the moving radio burst dominates the radio emission and its propagation direction is coincident with movement of the bright coronal front (dashed lines in Fig.~\ref{fig:nrh_source_temevo}). Then after a few minutes, the moving radio burst diminished into the background and the type IV emission started to dominate. The type IV emission occurred above the flaring region and moved outward in nearly solar radial direction. This propagation direction is coincident with the outward expansion of the associated CME.  In what follows, we concentrate primarily on the moving radio burst.

\begin{figure*}[!h]
\centering
\includegraphics[scale=1.]{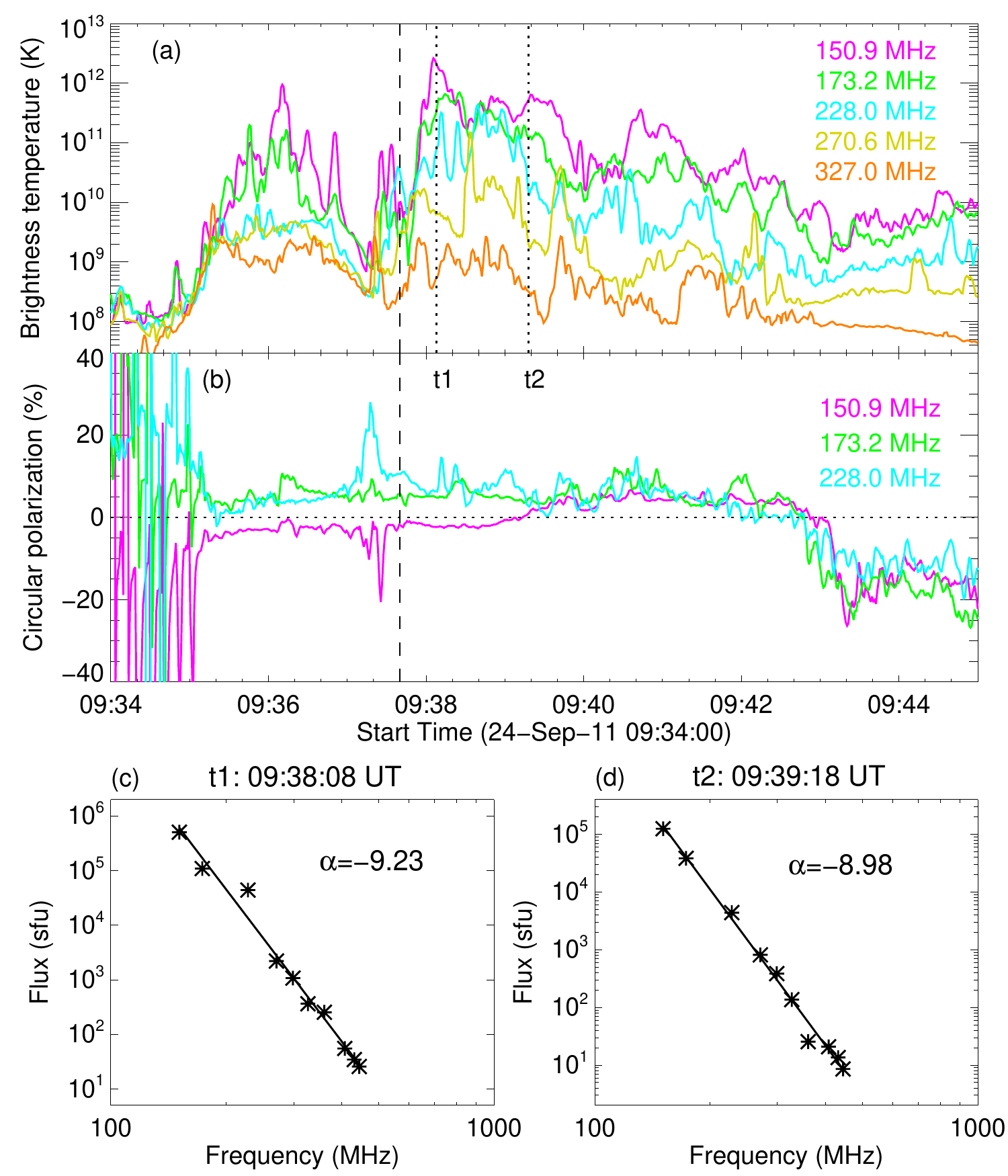}
\caption{Physical parameters of the moving radio burst. (a) The brightness temperature at five NRH frequencies (coded by different colors) between 150 and 327 MHz. (b) Degree of circular polarization at three NRH frequencies between 150 and 228 MHz. The vertical dashed line in each panel indicates the start time of the moving radio burst.
Panels (c) and (d) show the power-law fit to the NRH flux-frequency spectra at two different times, respectively. The spectral index $\alpha$ is shown in the upper-right corner of each panel.
\label{fig:local-lc}}
\end{figure*}

To determine the emission mechanism of the moving radio burst, the brightness temperature, the flux density, the circular polarization degree, as well as the spectral index of the moving radio burst are estimated (Fig.~\ref{fig:local-lc}).  
The brightness temperature is defined as an average values of all pixels within the 70\% maximum contour.
The flux density is computed for Stokes V and Stokes I components, respectively, based on which the circular polarization degree is calculated by taking the ratio between them (i.e., V/I). 
The spectral index ($\alpha$) is estimated by fitting a power-law function to NRH flux density spectra.
The vertical dashed line in Fig.~\ref{fig:local-lc} marks the start of the moving radio burst, after which the moving radio emission becomes strongly enhanced and consists of many bursty components (Fig.~\ref{fig:local-lc}a). 
Previous study has designated the moving radio burst as gyro-synchrotron emission radiated by energetic electrons trapped inside the CME \citep{James2017}. 
However, the spectral indices ($< -8$) calculated for two different times (t1 and t2) during the moving radio burst are both far beyond the known spectral index of gyro-synchrotron emission \citep{Bastian2001,Maia2007,Bain2014,Carley2017}.
This suggests that the moving radio emission does not necessarily come within the CME itself. 
Moreover, the brightness temperature at frequencies below 271 MHz are all above 10$^9$ K, especially that at 150.9 MHz, which even exceeds 10$^{12}$ K. 
Such a high temperature is indicative of a coherent emission mechanism such as the plasma emission or the electron cyclotron maser emission \citep{Dulk1985}. According to \cite{Melrose1975}, the maximum brightness temperature for the fundamental and the  second harmonic plasma emission from a gap distribution can be up to 10$^{16}$ K and 10$^{13}$ K, respectively.
However, throughout the moving radio burst, the radio emission between 150 and 228 MHz remains weakly polarized (circular polarization degree of $<$ 10\%, Fig.~\ref{fig:local-lc}b). 
The low degree of circular polarization strongly argues against the fundamental plasma emission and the electron cyclotron maser emission since both of them are supposed to be highly polarized ($\sim$ 100\%)\citep{Dulk1985}.
This leaves only the second harmonic plasma emission mechanism that can be used to interpret our observations. 


In the case of second harmonic plasma emission, the electron density ($n_e$) in the radio source region can be estimated from the observed emission frequency ($f_{obs}$) via the following:
\begin{eqnarray}
f_{obs}=2f_p  \\
f_{p}=8980\sqrt{n_e}
\end{eqnarray}
where $f_p$ represents the plasma frequency. Assuming that the radio emission originates from energetic electrons outside the CME,  the heliocentric distance and thus the 3D position of the radio source can be estimated if electron density models of the solar corona are provided.
Here we use the hybrid electron density model of \cite{Vrsnak2004} to estimate the heights of the moving radio source at specific frequencies. This model has taken into account the enhanced densities above the active regions and has the ability to smoothly connect the active region corona with the IP space.
The radio sources in 3D can then be projected  onto the STEREO-B/EUVI images to study their position relative to the EUV wave.
The enlarged dots and dashed circles in Fig.~\ref{fig:multiview}d represent the radio emission centroids and their  uncertainties ($\pm 0.2~R_{\odot}$)  at 150 MHz (magenta) and 228 MHz (cyan), respectively, and their projections onto STEREO-B/EUVI images are presented in Fig.~\ref{fig:multiview}c.
Note that the uncertainties are estimated by roughly taking half of the full width at half maximum of the radio emission.
As can be seen, the projected radio centroids locate right at the southwest flank of the EUV wave. The lower the frequency, the further away the source from the flank.

\begin{figure*}
\centering
\includegraphics[scale=.8]{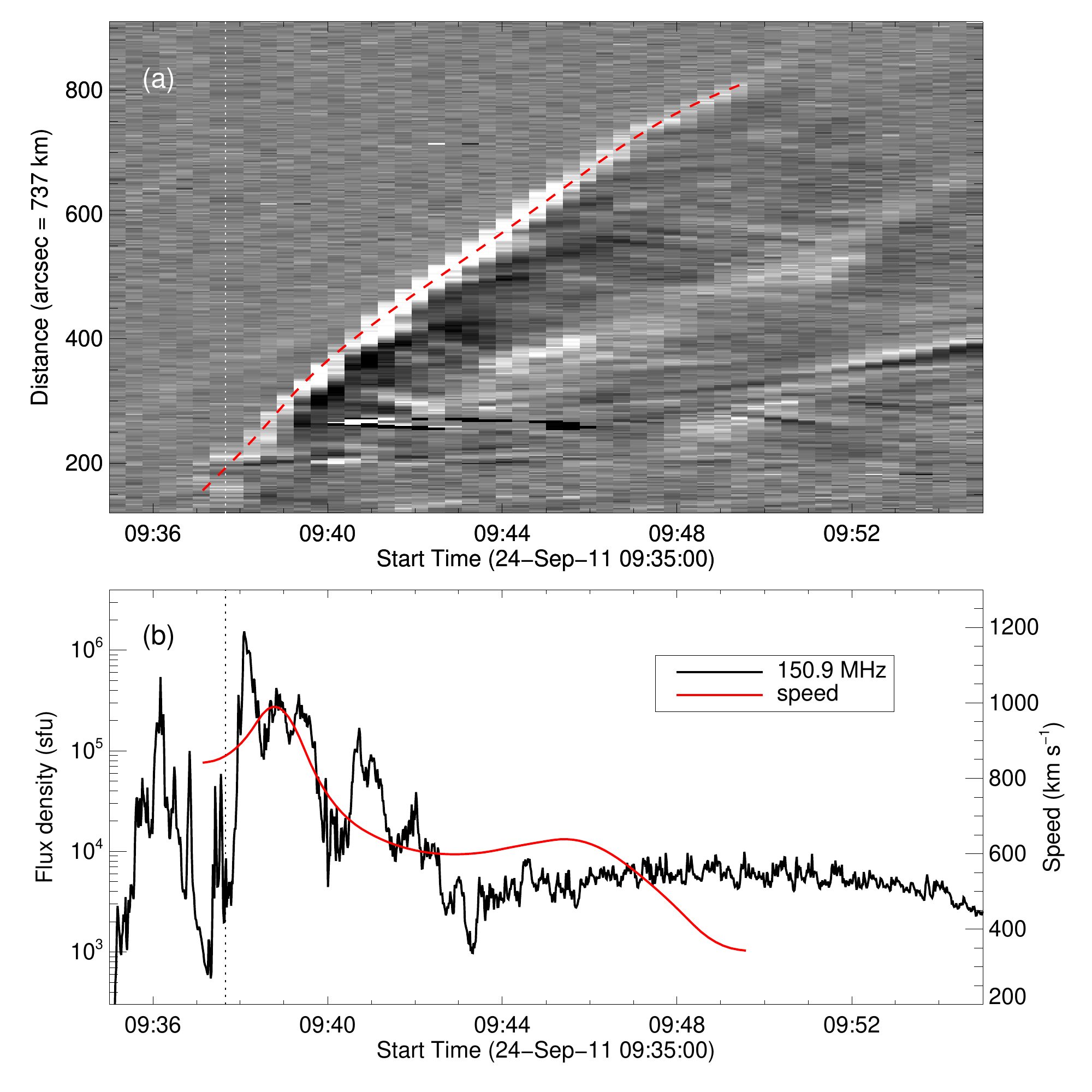}
\caption{Kinematics of the bright coronal front. (a) Running-difference space-time plot along the slit shown in Fig.~\ref{fig:bfront}f. The red dashed line represents the fifth degree polynomial fit to the bright front. The vertical dotted line shows the start of the moving radio emission. (b) Time profile of the disk-integrated radio flux density at 150.9 MHz (black line) and the POS speed of the bright front (red line). As can be seen, the radio flux density is well correlated with the propagation speed of the bright front. 
\label{fig:time_dis}}
\end{figure*}

To show the kinematics of the bright EUV front, a running ratio space-time plot sampled along a specific slit (white dashed line in Fig.~\ref{fig:multiview}f) in AIA 193 {\AA} images is presented in Fig.~\ref{fig:time_dis}a.
The vertical dotted line indicates the start of the moving radio burst, the red dashed line represents the fifth degree polynomial fit to the bright front.
The black line in Fig.~\ref{fig:time_dis}b shows the time profile of the disk-integrated flux density at 150.9 MHz, and the red line show the POS speed of the bright front. 
As can be seen, during the early period, the bright front was quickly accelerated to a speed of $\sim$ 1050 km s$^{-1}$, then decelerated to a nearly constant speed  ($\sim$ 600 km s$^{-1}$), then after about five minutes, the bright front decelerated again until it diminished into the thermal background.
Considering the projection effect, the true expansion speed of the corresponding EUV wave could be much larger than the values we calculated here. 
Moreover, the speed of the EUV wave was found to increase with its  heliocentric heights \citep{Cheng2012}.     
Of particular interest is the time evolution of the moving radio emission, which shows a high correlation with the propagation speed of the bright front, i.e., the faster the propagation, the stronger the radio emission (Fig.~\ref{fig:time_dis}b). 
Based on the density model of \cite{Vrsnak2004}, the heliocentric distance of the radio source at 150.9 MHz is estimated to be about 1.3 R$_\odot$. 
According to \cite{Zucca2014},  the local Alfv$\acute{\rm e}$n speed at such a height is about 700-1000 km s$^{-1}$, which is very close to the speed of the the bright front we show here.
If so, it is most likely that the EUV bright front (or the EUV wave) could be in fact a shock wave driven by the CME \citep{Mann2003}. In such a case, the electrons were accelerated by  the CME-driven shock and excite the local plasma to produce the second harmonic plasma  emission at radio frequencies. The bursty features in the radio emission imply that the electrons were accelerated in a quasiperiodic manner.
Given that the radio emission remains at low  altitudes in the solar atmosphere, it is likely that the electrons were accelerated at the flanks of the CME. This is different from the  electron acceleration at the CME nose, where type II radio emission usually emerges \citep{Su2016, Maguire2020}.

\section{Summary}

We have reported imaging and spectroscopy observations of the solar radio bursts on 24 September 2011. 
The dynamic spectrum shows multiple radio bursts, with a type III burst followed by a broadband continuum emission (type IV  burst). Interestingly, the type IV burst was superimposed with a much stronger moving radio burst.
The combination of high-cadence NRH (radio) and AIA (EUV) images has enabled us to locate the radio waves and discern where, when, and how the radio emission was produced.
Our observations show that the type III and type IV bursts originate from the solar flare and CME, respectively, while the moving burst is closely associated with a bright coronal front that propagated away from the active region.
According to the imaging observations from another perspective by STEREO-B/EUVI in 195 {\AA} (Fig.~\ref{fig:multiview}), the bright front is likely to correspond to the southwest flank of an EUV wave released from the active region. 

We then studied the physical parameter of the moving radio burst. The brightness temperatures at NRH frequencies below 271 MHz are all above 10$^9$ K, with the highest value even exceeding 10$^{12}$ K (Fig.~\ref{fig:local-lc}a). 
Meanwhile, the spectral indices calculated at two different times during the moving burst were found to be negative and steep ($<$ -8),  and the degree of circular polarization of the moving radio emission was found to remain at a very low value ($< 10\%$).
All above observational characteristics together with the bursty features observed in the moving radio emission suggest that the moving radio burst was produced via the second harmonic plasma emission mechanism.

Making using of the determined plasma emission mechanism and assuming the electron density model of \citep{Vrsnak2004}, the 3D position of the moving radio burst was estimated and projected onto the STEREO-B/EUVI images at 195 {\AA}. The projections show that the moving radio source locates right at the southwest flank of an EUV wave. The EUV wave expands rapidly both in radial and lateral directions, and is followed by a coronal dimming. 
A space-time plot along the propagation of the bright coronal front (Fig.~\ref{fig:time_dis}a) shows that the expansion speed of the EUV wave could be up to $\sim$ 1050 km s$^{-1}$. Considering projection effect,  its real speed could be  much larger than the local Alfv$\acute{\rm e}$n speed reported by \cite{Zucca2014}. This implies that the EUV wave may correspond to a shock wave driven by the CME. It was also found that the higher the EUV wave speed, the stronger the moving radio emission. We thus proposed that the moving radio burst results from energetic electrons accelerated by the CME-driven shock. These electrons excited the local plasma to oscillate and produce radio emission via the second harmonic plasma emission mechanism.

We perform this study using imaging observations from NRH and AIA. However, NRH only provides radio images at frequencies between 150 and 445 MHz, and AIA images were taken at a time cadence that is unable to match the fast time sampling of the radio images and spectra. 
In future, we would search more similar events that can be  observed by modern advanced radio imaging spectrometers such as  
 the Low-Frequency Array (LOFAR) \citep{Haarlem2013} and the Mingantu Ultrawide Spectral Radioheliograph (MUSER) \citep{Yan2016}. These radio telescopes can provide extremely high temporal and spectral resolutions in a very broad frequency range.
Moreover,  new EUV imagers such as the The Extreme Ultraviolet Imager (EUI) \citep{Rochus2020} aboard the Solar
Orbiter have been successfully launched into space, and the Ly$\alpha$ solar telescope (LST) \citep{Li2019,Feng2019} aboard the Advanced Space-based Solar Observatory (ASOS)  will be launched in the near future \citep{Gan2019,Gan2022}. 
Comparing to their precursors, these instruments have much improved temporal and spatial resolutions, and thus can have better synergy with the radio observations.
The combination of these advanced instruments would further improve our understanding of the solar radio burst.

\begin{acknowledgments}

We acknowledge the use of data from NRH, SDO/AIA, STEREO/EUVI, and the solar radio spectrometers of NDA, e-CALLISTO and WIND/WAVES. 
We would like to thank the referee for constructive suggestions.
This work is supported by NSFC (grant Nos. 12103090, U1731241, 11921003, 11973012, 11820101002), CAS Strategic Pioneer Program on Space Science (grant Nos. XDA15018300, XDA15052200, XDA15320103, and XDA15320301), the mobility program (M-0068) of the Sino-German Science Center, and the National Key R\&D Program of China (2018YFA0404200).
L.L. is also supported by CAS Key Laboratory of Solar Activity (KLSA202113).
 
\end{acknowledgments}

%
%
%
%

%
%


\bibliography{myref}{}
\bibliographystyle{aasjournal}



\end{document}